\documentclass[conference]{IEEEtran}
\IEEEoverridecommandlockouts
\usepackage{cite}
\usepackage{amsmath,amssymb,amsfonts}
\usepackage{algorithmic}
\usepackage{graphicx}
\usepackage{textcomp}
\usepackage{xcolor}
\usepackage{cite}
\usepackage[margin=10pt,font=small]{caption}
\usepackage{subcaption}
\usepackage{comment}
\usepackage[capitalise]{cleveref}
\usepackage{flushend}

\usepackage[top=54pt, left=54pt, right=54pt, bottom=54pt]{geometry}

\def\BibTeX{{\rm B\kern-.05em{\sc i\kern-.025em b}\kern-.08em
    T\kern-.1667em\lower.7ex\hbox{E}\kern-.125emX}}
\begin{document}

\title{\textcolor{black}{Haptic Teleoperation goes Wireless: Evaluation and \\Benchmarking of a High-Performance Low-Power Wireless Control Technology }}

\author{
\IEEEauthorblockN{Joseph Bolarinwa\(^*\), Alex Smith\(^*\), Adnan Aijaz\(^\dagger\), Aleksandar Stanoev\(^\dagger\), \\Mahesh Sooriyabandara\(^\dagger\), Manuel Giuliani\(^*\)  }
\IEEEauthorblockA{
\(^*\)\text{Bristol Robotics Laboratory, University of the West of England, Bristol, United Kingdom}\\
\(^\dagger\)\text{Bristol Research and Innovation Laboratory, Toshiba Europe Ltd., Bristol, United Kingdom}\\
firstname.lastname@brl.ac.uk; firstname.lastname@toshiba-bril.com}
} 


\maketitle

\begin{abstract}
Communication delays and packet losses are commonly investigated issues in the area of robotic teleoperation. This paper investigates  application of a novel low-power wireless control technology (GALLOP) in a haptic teleoperation scenario developed to aid in nuclear decommissioning. The new wireless control protocol, which is based on an off-the-shelf Bluetooth chipset, is compared against standard implementations of wired and wireless TCP/IP data transport. Results, through objective and subjective data, show that GALLOP can be a reasonable substitute for a wired TCP/IP connection, and performs better than a standard wireless TCP/IP method based on Wi-Fi connectivity. 
\end{abstract}


\section{Introduction}
Robot teleoperation, which involves the remote manipulation of robotic systems, continues to find application in different fields of robotics including industrial robots \cite{Gonzalez2021}, mobile ground robots \cite{Opiyo2021,Kot2018}, assistive robots \cite{Bolarinwa2019}, medical and surgical robots \cite{Feizi2021} \cite{Yang2020}, robots for nuclear environments \cite{Qian2012,Aleotti2017}, and telepresence robots\cite{Niemela2019}. Depending on the application, the leader and follower robots (that form the teleoperation system) may be situated in the same environment (with physical obstruction between the teleoperator and the robot) \cite{Dogangil2009,Arata2008,Zemiti2004}, or thousands of miles away from each other \cite{Avgousti2016}. In robot teleoperation, the teleoperator sends position and control commands and in turn receives visual and other sensory feedback information from the remote end. 

 Applications that involve the teleoperation of a robot within the same local environment as the operator\cite{Dogangil2009} may allow for wired connectivity between the teleoperated robot and the operators' control base. However, other applications may require that the operator controls a robot many miles away, or may encounter limitations to wired connectivity such as limited cable length or cable disconnect. In such cases, wireless robot teleoperation may be explored.

One of such applications where wireless robot teleoperation is currently being explored is in nuclear decommissioning, due to the health implications of exposure to nuclear radiations. However, the complexities of nuclear facilities and the risk of cable disconnect create doubts on the sole use of wired connectivity, hence the need to explore wireless robot teleoperation. Due to the real-time control-data communication requirements of robot teleoperation, wireless communication systems replacing the existing wired communication systems should have similar or better performance, particularly in terms of end-to-end delay/latency, jitter (latency variations), and congestion.  

To this end, the main objective of this paper is to report the key differences in performance between wired and wireless communication protocols. We present our design and implementation of the TCP/IP protocol for wired and wireless robot teleoperation and an implementation of the GALLOP wireless control protocol \cite{GALLOP_journal} for robot teleoperation - in particular we aim to show that the novel GALLOP protocol can be considered as a wireless alternative to a wired connection without loss of control quality. We compared how long it takes to send and receive data between the leader and follower robots, as well as the position and velocity errors for all the protocols examined. We also report on a heuristic evaluation of the wired and wireless robot teleoperation. In the heuristic evaluation, we examined how responsive the follower robot is to changes at the leader robot, the smoothness of the control, and the safety of the control. For clarity, wired TCP/IP and wireless TCP/IP (based on standard Wi-Fi) are  simply referred to as wired/wireless respectively through the rest of this paper.


\textcolor{black}{It is emphasized that low-power wireless technologies are predominantly used for monitoring and non-critical applications. To the best of our knowledge, this is one of the first works employing a low-power wireless control technology (implemented on an off-the-shelf Bluetooth chipset) for haptic teleoperation in a real-world test bed. }

\section{Related Work}
\subsection{Wireless Technologies for Teleoperation}




Wireless communication technologies have evolved at an unprecedented pace over the past three decades. 
Wireless technologies have also been used to transmit haptic information (kinesthetic and tactile) between leader robots and follower robots. Bilateral haptic communication implies that the interactions between the follower robot and the remote environment reflect back to the operator, hence influencing how the operator reacts. In order to enable real-time interactions as well as to provide system stability and transparency, bilateral haptic feedback control loops for the leader and follower robots impose a 1 kHz frequency update rate \cite{Oparin2017}. It is therefore vital to consider how network conditions might affect haptic applications when choosing communication protocols for teleoperation in order to provide high quality of experience (QoE). Different protocols have been developed with respect to the Internet protocol suite networking model \cite{Kokkonis2012}. 


TCP and UDP protocols are the most commonly used transport layer protocols for haptic communication to demonstrate physical interactions between human operators and remote environments or for communication between physical devices and virtual environments. Available protocols can also be classified based on parameters like network delay, jitter, packet loss, and rate of data transfer. 

 Application layer protocols for haptic communication enable aggregation and multiplexing of audio, video and haptic data streams. This is important because quality of experience requirements demand synchronized transmission of video and audio data, with real-time haptic interaction between the user and the remote environment. Some examples of application protocols include Session Initiation Protocol (SIP) \cite{King2010}, Real-Time Protocol (RTP) for distributed interactive media(RTP/I) \cite{Mauve2001}, and Application Layer Protocol for Haptic Networking (ALPHAN) \cite{Osman2007}. For application in telesurgery, \cite{King2009} present an application layer protocol, referred to as the Interoperable telesurgery Protocol (ITP). However, communication using this protocol was not bilateral as it was used to transmit only video data. The Haptics over Internet Protocol (HoIP) uses UDP and a multiplexing algorithm which enables packetization audio/haptic or video/haptic data \cite{Gokhale2013HoIP}. 
 


\textcolor{black}{Recently, fifth-generation (5G) mobile/cellular technology has received significant attention for haptic teleoperation due to native support for ultra-reliable low-latency communication (uRLLC). The requirements and design challenges for haptic communication over 5G have been identified in \cite{aijaz_wcm}, \cite{TI_5G}, and \cite{aijaz_PIEEE}. Wireless resource allocation enhancements for meeting the requirements of haptic communication over 5G have been investigated in \cite{aijaz_twc} and \cite{aijaz_sj}. Although 5G is promising for haptic teleoperation, guaranteed latency and timeliness for packet delivery is not possible without scheduling enhancements. Besides, real-world trials of haptic teleoperation over 5G are in infancy. }



\subsection{Stability Control Architectures for Teleoperation}\label{sec:stabTele}

Long distance communication introduces varying amount of latency which makes certain applications of teleoperation difficult and/or impossible due to instability. There are however stability control architectures and methods that are employed to minimise the impact of latency on the stability of applications like teleoperation. The introduction of adaptive control subsystems also has advantages and disadvantages, and the choice of which control scheme to employ depends on applications. Comparison of different control schemes was carried out by \cite{Arcara2002}. Classifying bilateral control of teleoperation systems can be based on the choice of either compensating for communication delays, estimation of the operator and environment model, handling of internal and external disturbances of the subsystems, or a combination of the highlighted tasks. 

Wave variable control, time-domain passivity approach, and  model-mediated tele-operation are some of the key available control schemes that address stability and communication challenges for networked teleoperation systems. Using algorithms created to ensure stability and transparency between leader and follower devices when time delay is introduced, \cite{Niemeyer1991,Niemeyer1998} conceptualised the wave-variable control method. It builds on the work of \cite{Anderson1988} which combines scattering transformation, network theory and passive control. 
The time-domain passivity control (TDPC) \cite{Ryu2010} monitors the energy flowing to and from the leader side, follower side, or both in real time by using a passivity observer (PO) placed in series or parallel to the communication channel. In the TDPC, a passivity controller (PC) retains the system's passivity through the use of adjustable damping elements. In order to ensure system stability and transparency in the presence of arbitrary communication delay, the model-mediated tele-operation approach (MMTA) was proposed \cite{achhammer2010improvement}. Instead of directly sending back haptic (force) signals, parameters of the object model (which approximates the remote environment) are estimated and transmitted back to the master in real-time as the slave interacts with the remote environment. 

\textcolor{black}{\section{Overview of GALLOP Technology}}
\textcolor{black}{This work employs a high-performance wireless control technology, i.e., GALLOP, as a wire-replacement technology for haptic teleoperation. GALLOP has been designed for wireless closed-loop control or feedback control in single-hop as well as multi-hop scenarios. GALLOP is capable of handling control loops with ultra-fast dynamics on the order of milliseconds (ms). GALLOP implements a control-aware bi-directional schedule that handles cyclic exchange of control information with very low latency and zero jitter. GALLOP also implements various techniques for achieving very high reliability in harsh wireless environment. GALLOP is agnostic to the Physical (PHY) layer design; hence it can be implemented on different wireless chipsets including those of Bluetooth and Wi-Fi. Further technical details about GALLOP are available in \cite{Aijaz2019} and \cite{GALLOP_journal}. In our work, GALLOP provides wireless connectivity for bi-directional haptic data exchange between the leader and the follower robot. We realize this communication based on a Bluetooth 5.0 wireless chipset.}

\begin{figure}[htbp]
\centering
  \includegraphics[width=0.8\columnwidth,clip]{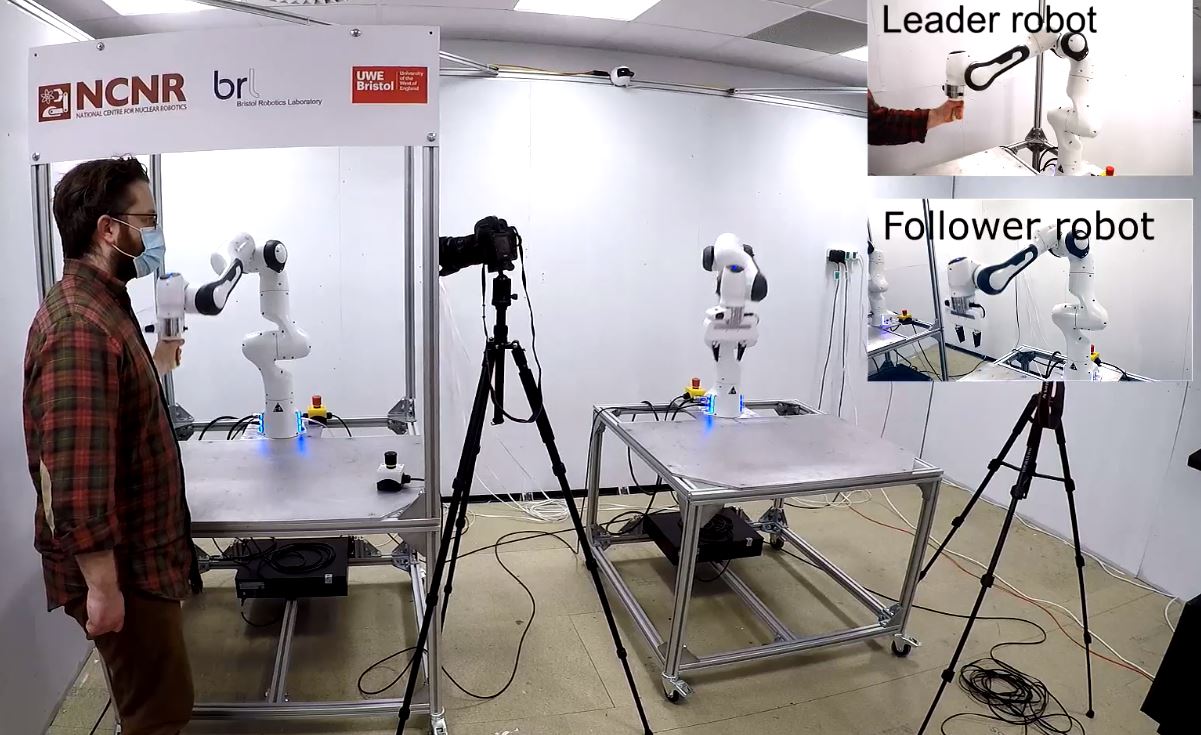}

\caption{Setup for leader and follower robots.}
\label{fig:leader_and_follower_robot_demonstration}
\end{figure}

\section{Teleoperation Setup}
In this section, we describe the hardware and software setup for the three scenarios explored in this study. The study was carried out at the Nuclear Robotics test bed of the Bristol Robotics Laboratory, Bristol. The teleoperation setup comprises of two sets of robotic manipulators. The first is the leader robot, at the operator end, where commands are issued. The second robot is the follower, designed to replicate the movements of the leader robot, hence the reference ``leader-follower''. \cref{fig:leader_and_follower_robot_demonstration} shows the leader robot and follower robot setup with an operator demonstrating the process. 
The robot used on both end is the Franka Emika Panda robot arm \cite{FrankaEmika2021}. Computations on and communication between the leader and follower robots is carried out using an Nvidia Jetson Xavier board \cite{Nvidia2021} connected to the controller of each arm. Based on the real time control loop requirement for efficient teleoperation, Ubuntu operating system with real-time kernel was installed on the Jetson boards. Programs written to implement data transfer and processing were written in C++ and run on the Jetson boards connected to each robot controller.  

\begin{figure}[htbp]
\centerline{\includegraphics[width=0.8 \columnwidth,clip]{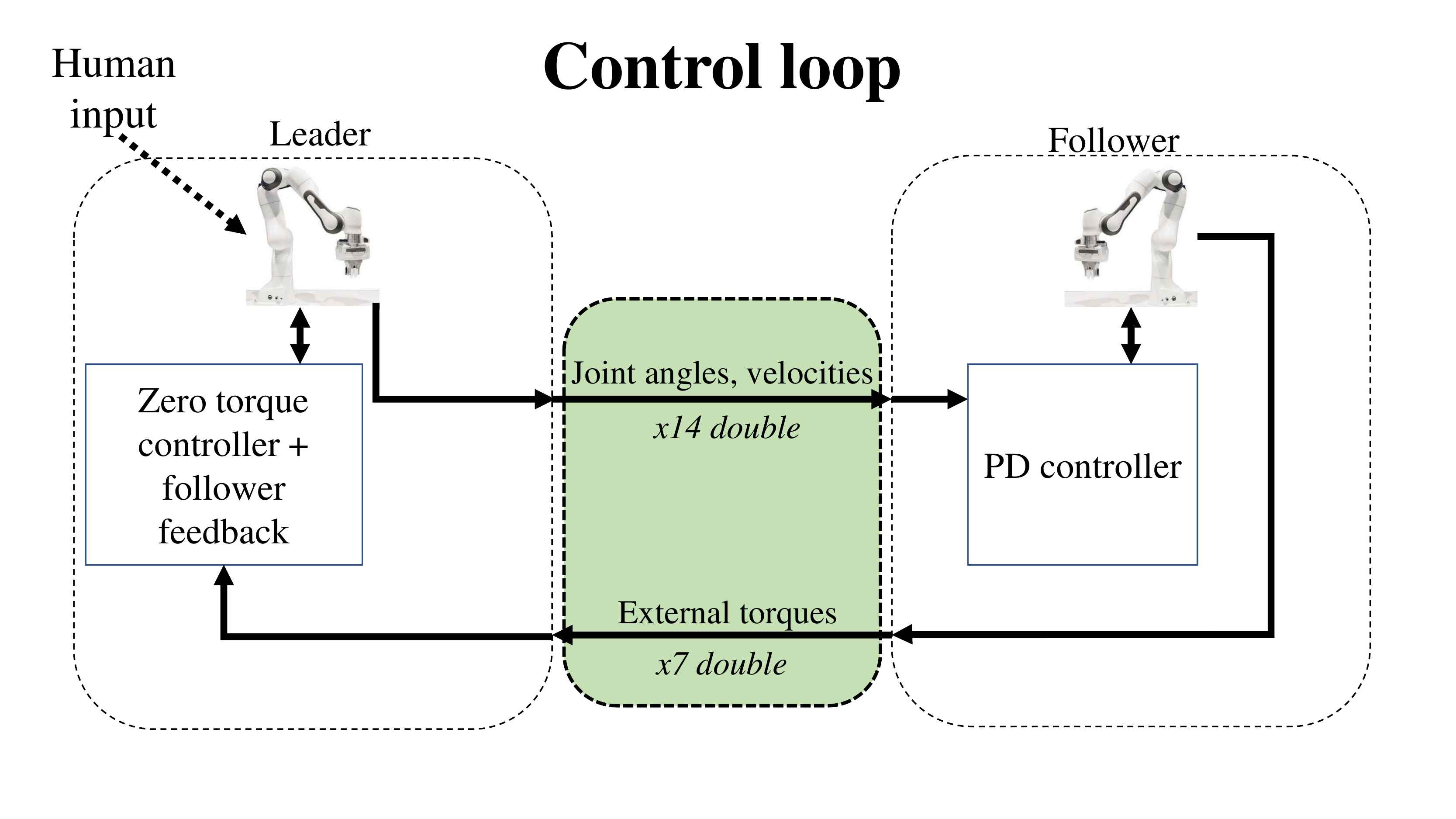}}
\caption{Control loop for leader and follower robots.}
\label{fig:control loop}
\end{figure}

The control loop for the leader and follower robots is shown in \cref{fig:control loop}. As the leader robot is moved, joint angles and velocities of the leader robot are sent to the follower robot (which replicates the leader robot's movements) at the remote end. Simultaneously, external torques on the follower robot are sent back to the leader robot moved by the operator. The command torque for the leader side is defined as:
\begin{equation}
    \tau_d^L(t) = K \tau_{ext}^F(t), \quad K < 1
    \label{eq:tauL}
\end{equation}
where $\tau_d^L \in \mathbb{R}^7$ is the desired torque for the leader arm, $K  \in \mathbb{R}$ is a scaling factor and $\tau_{ext}^F \in \mathbb{R}^7$ is the measured external torques being applied at the follower side. For the follower robot control 
\begin{equation}
    \tau_d^F(t) = P\left(q^F\left(t\right) - q^L\left(t\right)\right) - D\left(\dot q^F\left(t\right) -\dot q^L\left(t\right)\right)
    \label{eq:tauF}
\end{equation}
where $\tau_d^F \in \mathbb{R}^7$ is the desired follower-side torque, $P$ and $D$ are diagonal gain matrices, $q^F, q^L \in \mathbb{R}^7$ are follower and leader joint angles respectively, and $\dot q^F, \dot q^L \in \mathbb{R}^7$ are follower and leader joint velocities. Data transfer is therefore limited to a single 7-double vector in \cref{eq:tauL} and two 7-double vectors in \cref{eq:tauF}. The control loop for each robot runs at 1kHz, with the communication loop running in a separate thread at 20Hz.   
\begin{figure}[htbp]
\centering
\subfloat[]{%
  \includegraphics[width=0.85\columnwidth,clip]{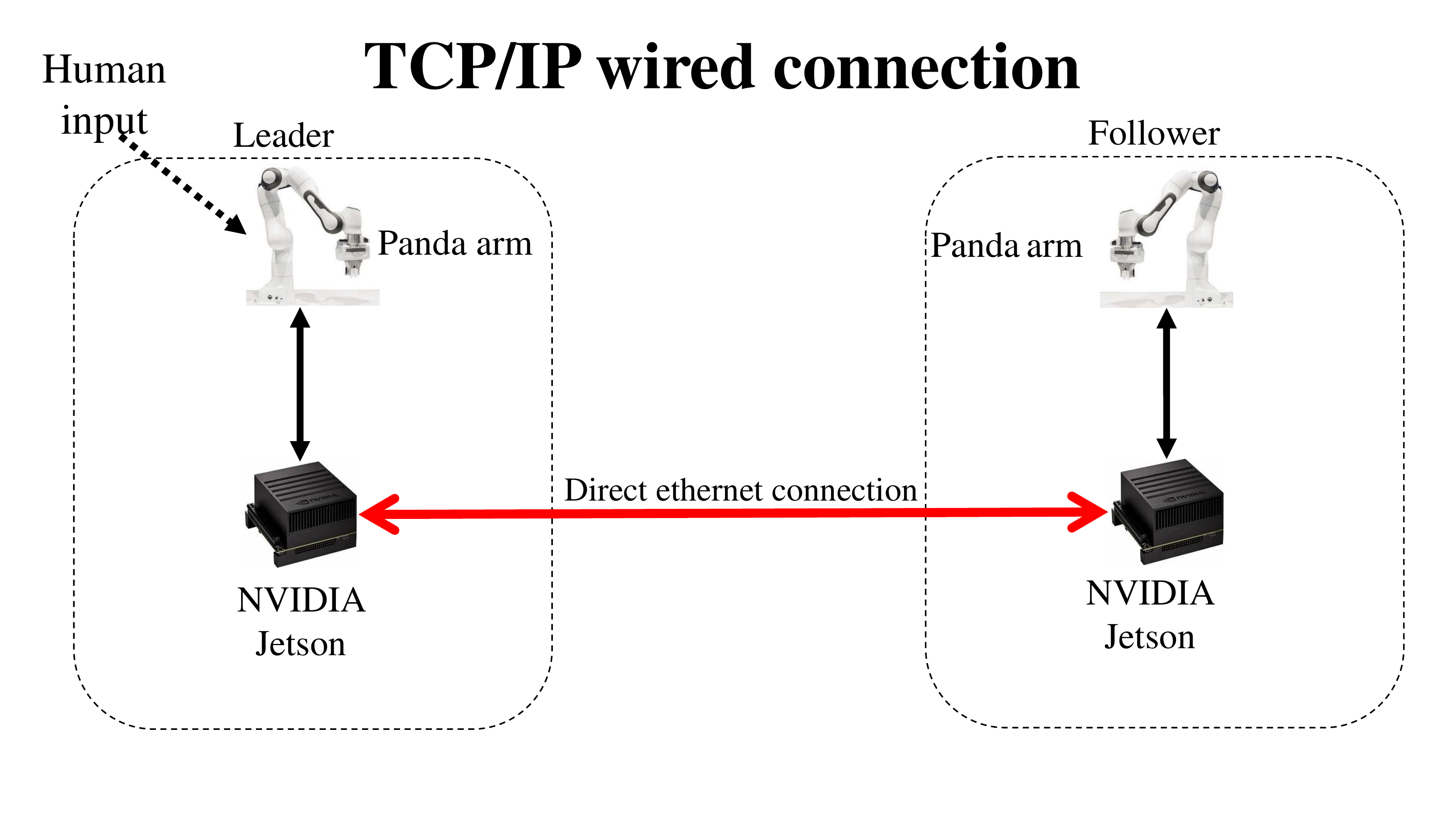}\label{fig:tele-operation_setup:wired_tcpip}%
}

\subfloat[]{%
  \includegraphics[width=0.85\columnwidth,clip]{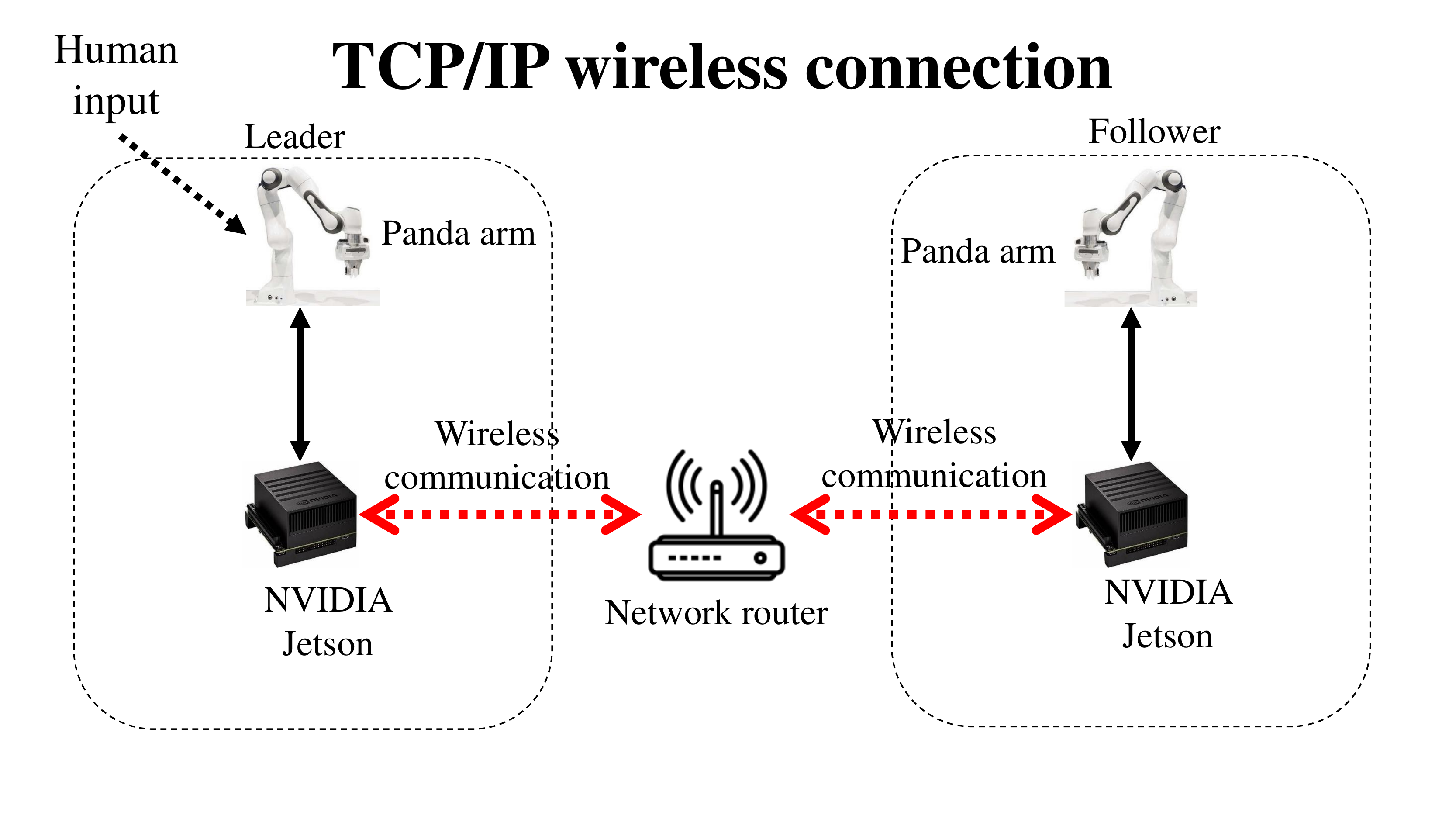}\label{fig:tele-operation_setup:wireless_tcpip}%
}

\subfloat[]{%
  \includegraphics[width=0.85\columnwidth,clip]{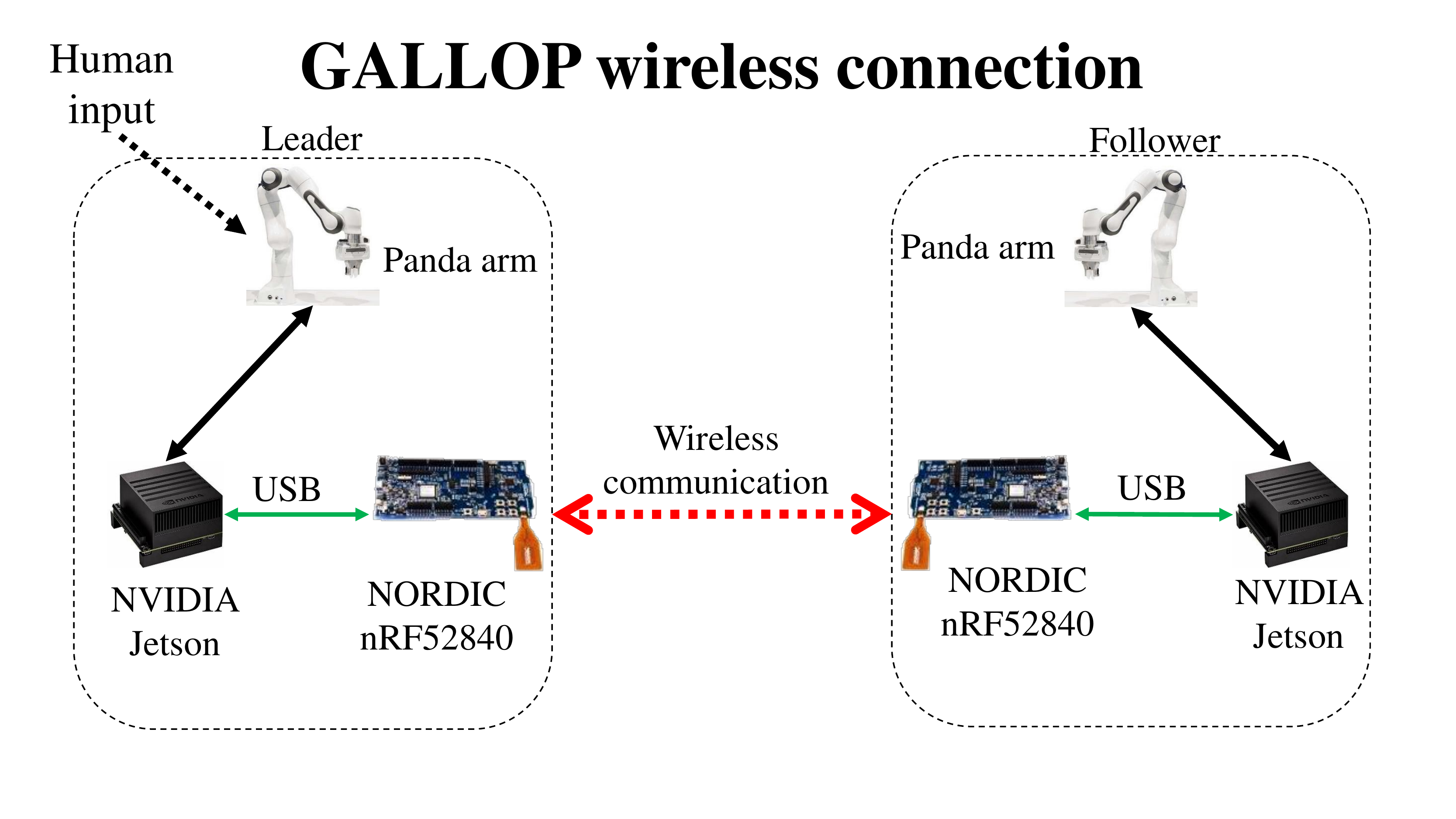}\label{fig:tele-operation_setup:wireless_gallop}%
}

\caption{Robot teleoperation setup (a) wired, (b) wireless, and (c) GALLOP-based.}
\label{fig:tele-operation_setup}
\end{figure}
In \cref{fig:tele-operation_setup:wired_tcpip}, the physical wired connection between the leader robot and follower robot is shown. Whilst the leader and follower Jetson boards were physically connected using a LAN cable (wired connection), programs for running the TCP/IP communication protocol on either side were run on the Jetson Xavier boards. 

As shown in \cref{fig:tele-operation_setup:wireless_tcpip}, the only difference between the wired connection and the wireless connection is that the physical wired LAN connection was replaced with a network router for wireless communication between the leader and follower robots.  

In \cref{fig:tele-operation_setup:wireless_gallop}, we introduce the GALLOP protocol to allow wireless communication of haptic data between the leader and follower robots. The GALLOP files were uploaded onto NORDIC nRF52840 boards \cite{NordicSemiconductors2021}, connected to the Jetson boards responsible for processing data transmission to and from each robot.   

\section{Evaluation}
A heuristic evaluation of the three communication protocols of interest was carried out. Heuristic evaluations are often employed in the field of human-computer-interface (HCI) as part of the design cycle as a usability inspection method. Heuristic evaluations require that experts use their practical skills in combination with theoretical knowledge of standards and guidelines \cite{Yeratziotis2017}. During the evaluations, experts carry out tasks against previously determined usability principles referred to as heuristics that when violated make the system more difficult to use \cite{Murtza2017}. Heuristic evaluations have previously been carried out on smartphone applications in supporting elderly \cite{Salman2018}, virtual reality systems \cite{Murtza2017}, design and development of a statistics serious game \cite{Bunt2017}, and mobile applications \cite{Joyce2018}. 

In the study reported in this paper, five robotics experts (mean years of robot experience = 10.8 years) carried out heuristic evaluations of the three communication protocols (wired, wireless, GALLOP) as they carried out a task of sorting six objects into three containers (two objects per container). During the heuristic evaluation, the leader and follower robots were placed in the same room, so the experts were able to see the movements of the follower robot in real time. For each experiment run, the objects to be sorted were placed randomly in front of the follower robot. The expert participants moved the leader robot by hand and used a keypad button to open/close the robot's gripper. The experts carried out three repetitions of the task for each of the communication protocols in a randomised order.   

For each task scenario, we measured the send and receive times of packets sent from the leader robot and follower robot. Position and velocity errors were also measured for each communication scenario explored. At the end of each task, the expert participants completed a heuristic questionnaire, which consisted of three 5-point Likert scales rating the responsive, feeling of safety, and feeling of smoothness of the robot control.  

\section{Results}
To compare the performance of the different communication methods objectively, we calculate errors from position and velocity in time for each experimental run, then calculate error indexes $\epsilon$ and $\dot \epsilon$ from the root-mean-square (RMS) of errors. 
\begin{equation}
    e(t) = q_l(t) - q_f(t)
\end{equation}
Where $e(t) \in \mathbb{R}^7$ denotes joint position error at time $(t)$, $q_l(t) \in \mathbb{R}^7$ is leader manipulator joint angles and $q_f(t) \in \mathbb{R}^7$ is follower manipulator joint angles. Similarly, for velocity error $\dot e(t) \in \mathbb{R}^7$:
\begin{equation}
    \dot e(t) = \dot q_l(t) - \dot q_f(t)
\end{equation}
where $\dot q_l(t) \in \mathbb{R}^7$ is leader manipulator joint velocities and $q_f(t) \in \mathbb{R}^7$ is follower manipulator joint velocities. From these error values we can generate Root Mean Square (RMS) values for each joint $1 \leq i \leq 7$ as 
\begin{align}
    rms_i &= \sqrt{\frac{1}{T} \sum_{t=0}^{T} |e_i(t)|^2}, \notag \\
    \dot{rms}_i &= \sqrt{\frac{1}{T} \sum_{t=0}^{T} |\dot{e}_i(t)|^2}
\end{align}
where $rms$ and $\dot{rms} \in \mathbb{R}^7$. These values at this point can be used to generate an error score for each experimental run by summing the RMS over all joints which will be represented by the variables $\epsilon$ and $\dot{\epsilon}$ for RMS of position and velocity error respectively: 
\begin{align}
    \epsilon &= \sum_{i=1}^{n} rms_i, \notag \\
    \dot{\epsilon} &= \sum_{i=1}^{n} \dot{rms}_i
\end{align}

\begin{figure}[htbp]
\centerline{\includegraphics[width=0.75\columnwidth]{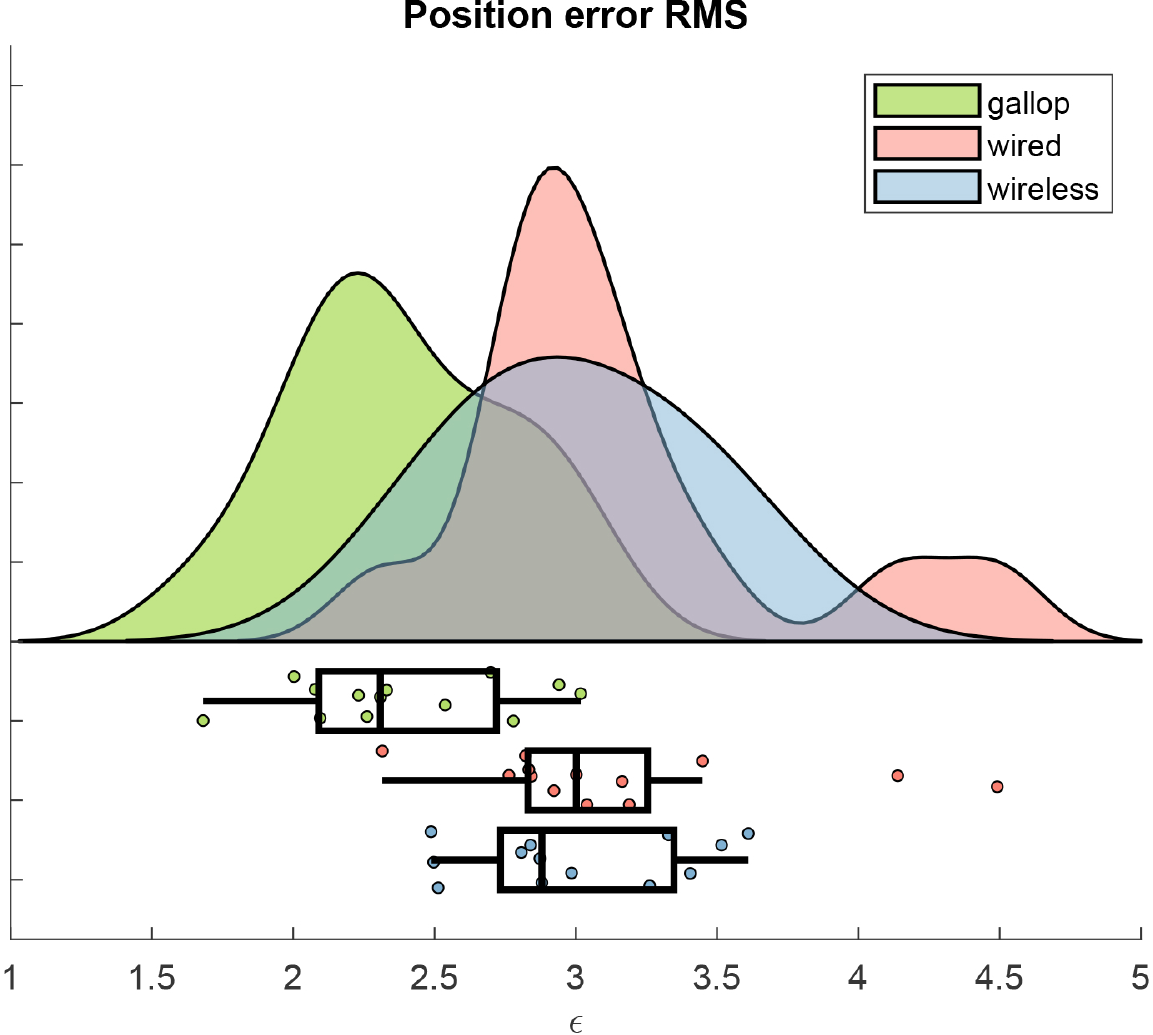}}
\caption{Distributions of $\epsilon$ for the three communication methods.}
\label{fig:eRMS}
\end{figure}

\begin{figure}[htbp]
\centerline{\includegraphics[width=0.75\columnwidth]{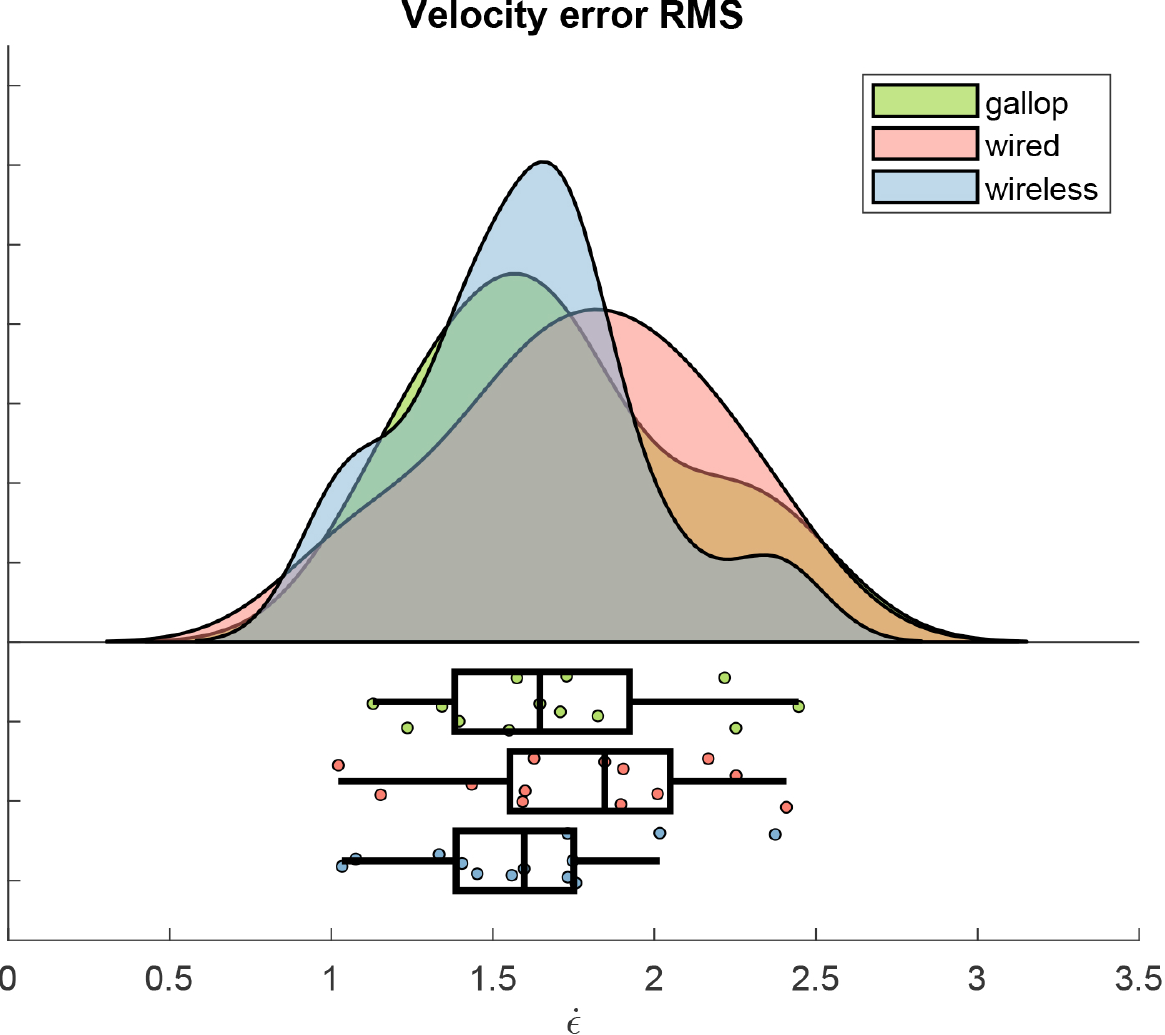}}
\caption{Distributions of $\dot\epsilon$ for the three communication methods.}
\label{fig:edotRMS}
\end{figure}

\begin{figure}[htbp]
\centering
\subfloat[]{%
  \includegraphics[width=0.8\columnwidth]{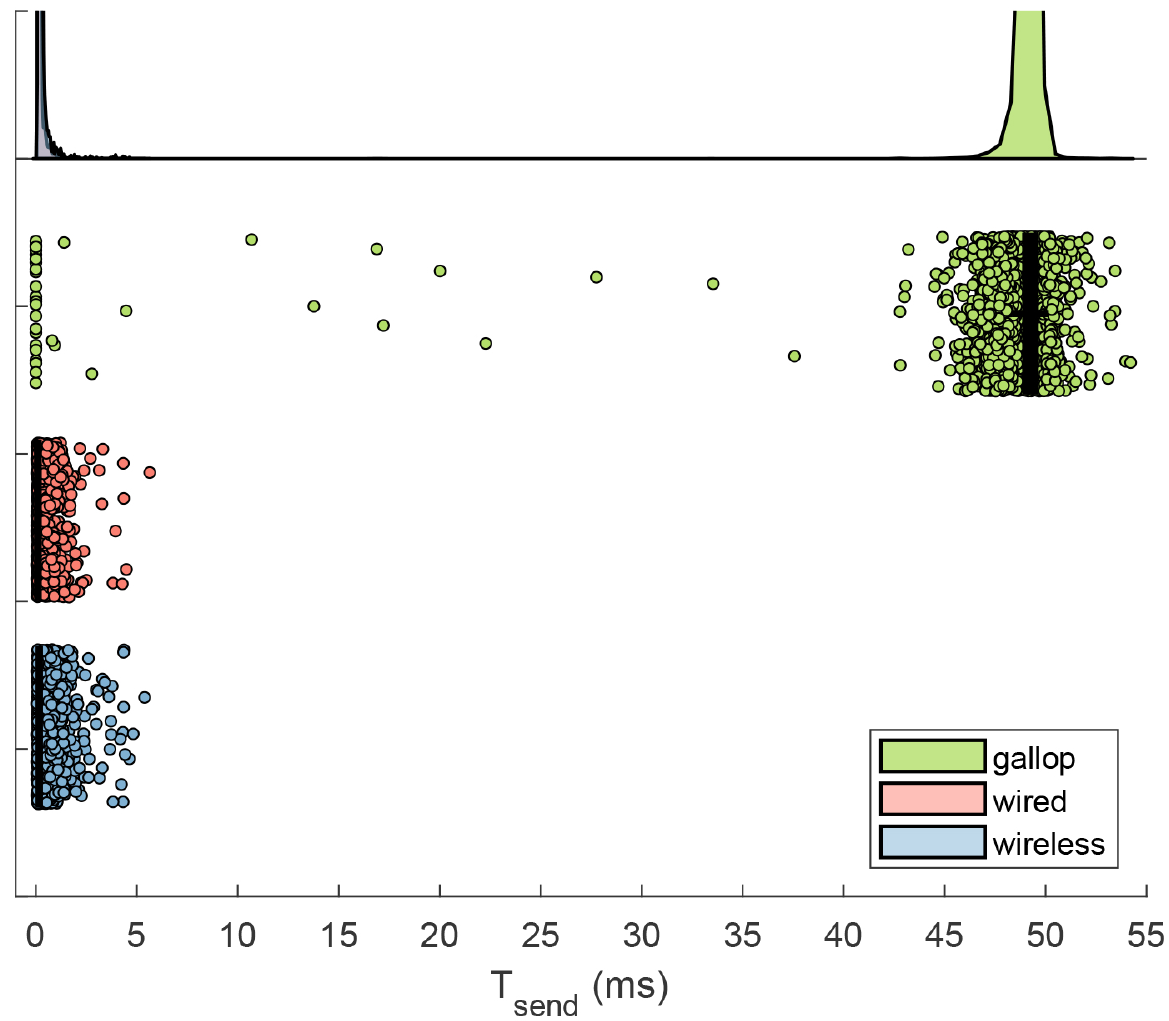}
  \label{fig:leader_times:send}%
}

\subfloat[]{%
  \includegraphics[width=0.8\columnwidth]{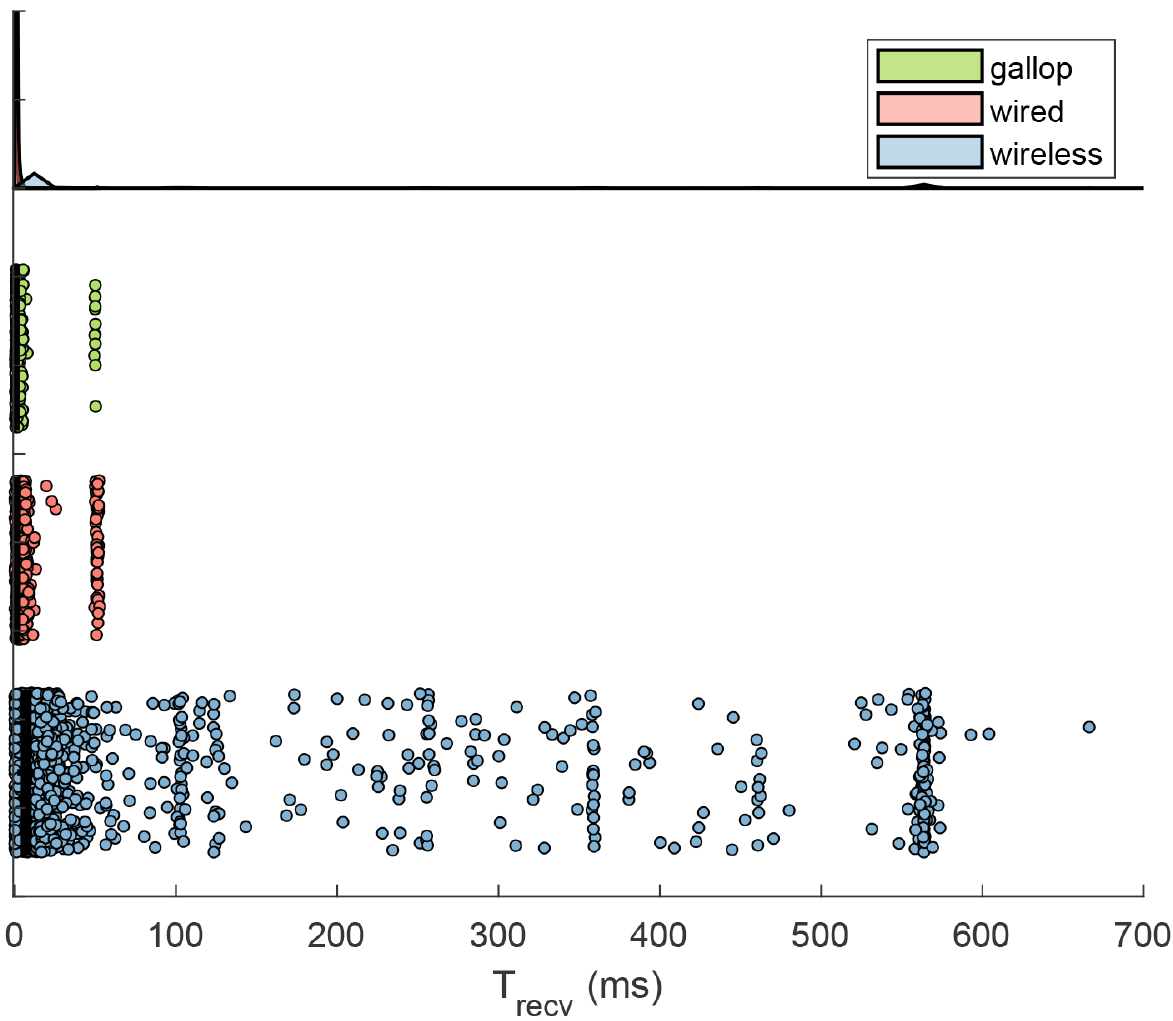}
  \label{fig:leader_times:recv}%
}
\caption{Leader-side data transmit times (a) and receive times (b), for the three communication methods. Follower-side distributions were found to be similar, so are not presented here for brevity.}
\label{fig:leader_times}
\end{figure}

\begin{figure}[htbp]
\centerline{\includegraphics[width=0.75\columnwidth]{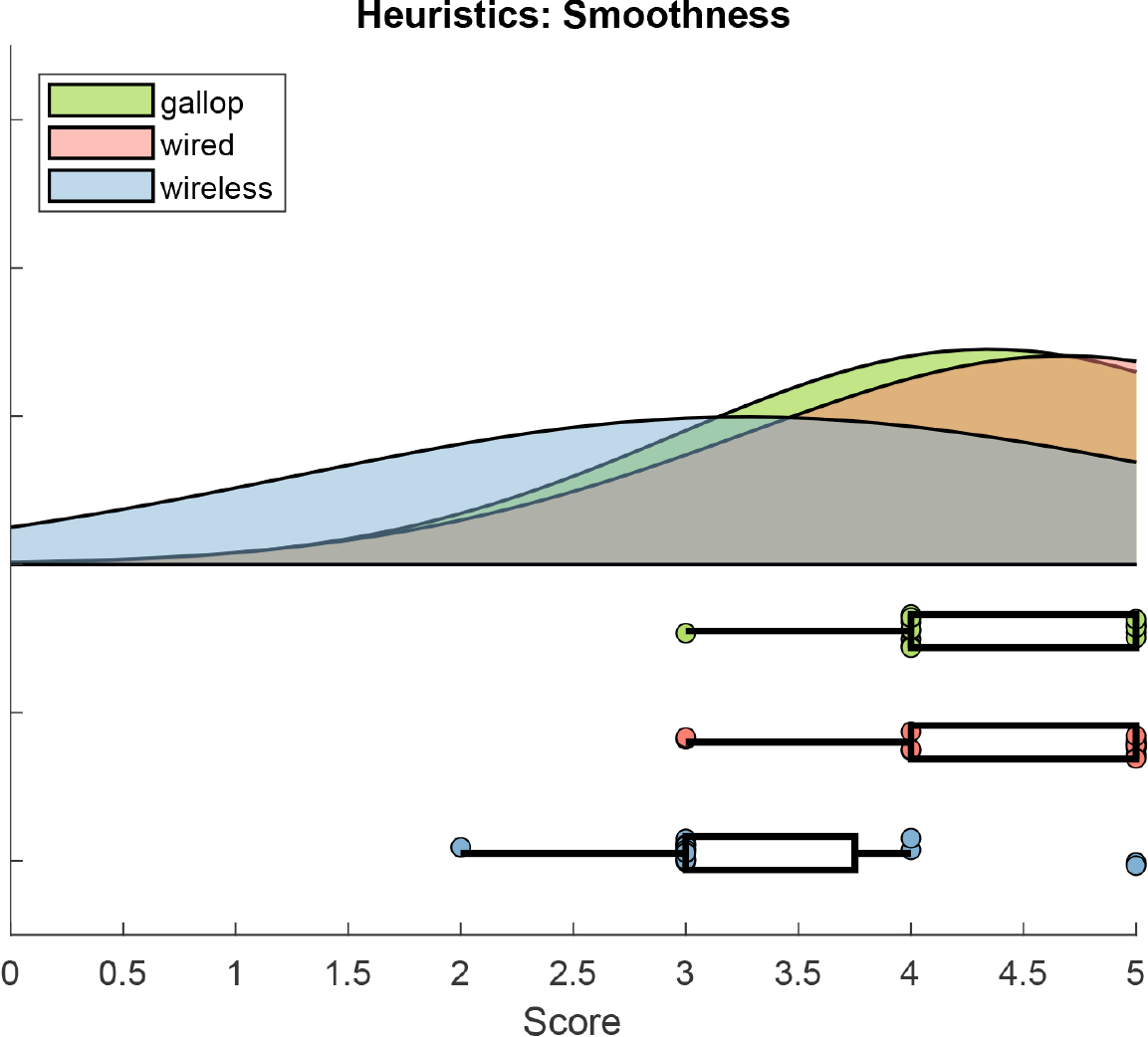}}
\caption{Scores for perceived "smoothness".}
\label{fig:smoothness}
\end{figure}

\begin{figure}[htbp]
\centerline{\includegraphics[width=0.75\columnwidth]{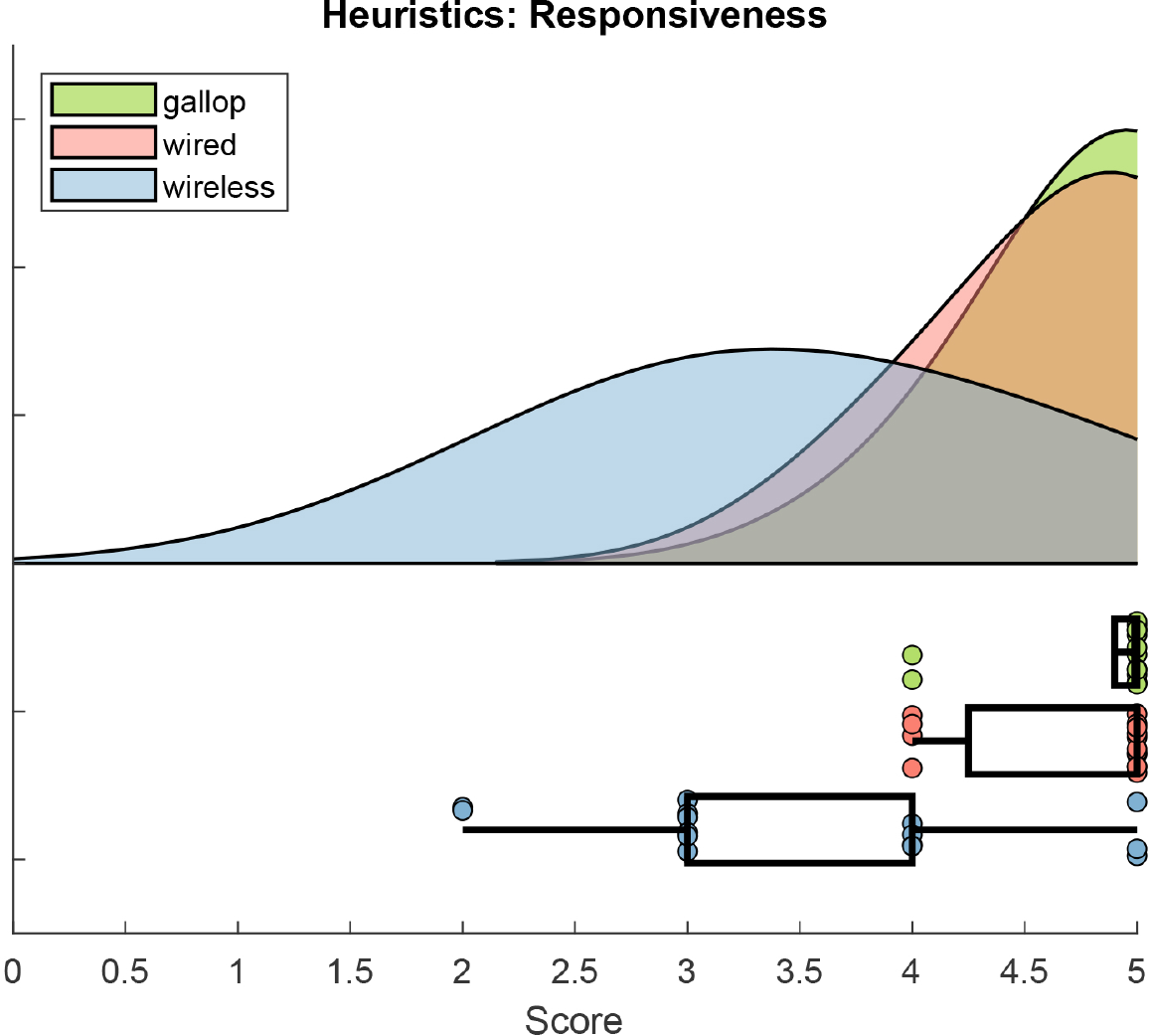}}
\caption{Scores for perceived "responsiveness".}
\label{fig:responsive}
\end{figure}

\begin{figure}[htbp]
\centerline{\includegraphics[width=0.75\columnwidth]{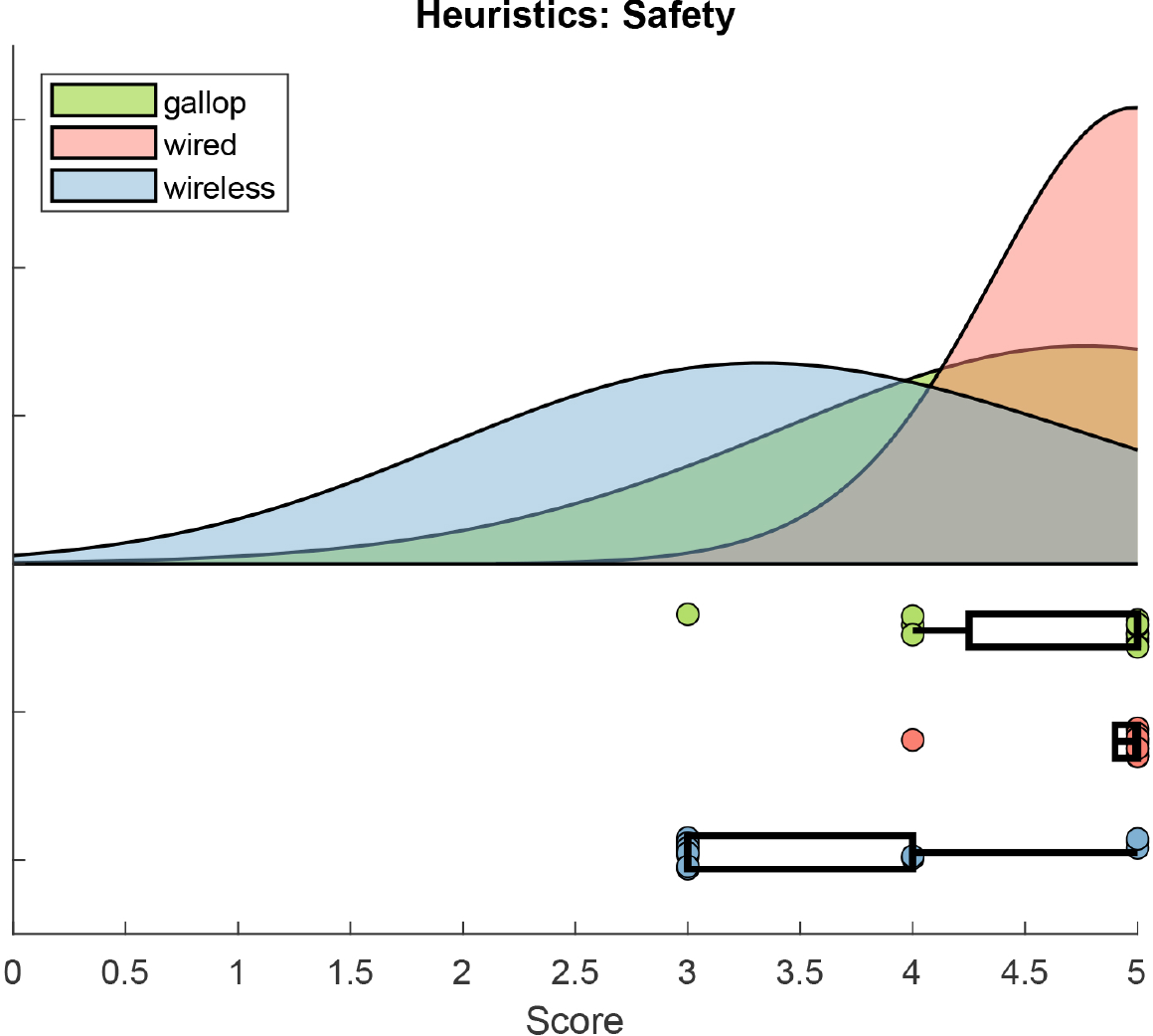}}
\caption{Scores for perceived safety.}
\label{fig:safe}
\end{figure}

\begin{table}[htbp]
\scriptsize
\centering
\caption{Statistics for $T_{send}$}
\begin{tabular}{l|r|r|r|r|r|r}
            & \multicolumn{2}{|c|}{GALLOP} & \multicolumn{2}{|c|}{wired} & \multicolumn{2}{|c}{wireless} \\
            & Lead        & Follow        & Lead          & Follow        & Lead          & Follow \\
            \hline
N           & 35,478      & 35,635        & 30,359        & 60,464        & 27,636        & 55,040 \\
Mean (ms)   & 49.1      & 49.1        & \textbf{0.116} & 0.126         & 0.178         & 0.217 \\
$\sigma$    & 1.61       & 1.51        & 0.127         & \textbf{0.093}  & 0.176        & 0.153  \\
Range (ms)  & 54.2      & 63.3       & 5.59         & \textbf{3.45}  & 5.36         & 8.91   \\
IQR (ms)    & 0.451      & 0.541         & \textbf{0.039} & 0.046       & 0.049        & 0.061 
\end{tabular}
\label{tab:tsend}
\end{table}

\begin{table}[htbp]
\scriptsize
\centering
\caption{Statistics for $T_{recv}$}
\begin{tabular}{l|r|r|r|r|r|r}
            & \multicolumn{2}{|c|}{GALLOP} & \multicolumn{2}{|c|}{wired} & \multicolumn{2}{|c}{wireless} \\
            & Lead        & Follow        & Lead          & Follow        & Lead          & Follow \\
            \hline
N           & 35,478      & 35,635        & 30,359        & 60,464        & 27,636        & 55,040 \\
Mean (ms)   & \textbf{0.688} & 0.786        & 1.26         & 0.994         & 12.9         & 12.7 \\
$\sigma$    & 1.03       & \textbf{0.622}   & 2.37         & 1.88            & 51.7        & 52.8  \\
Range (ms)  & 50.1      & \textbf{35.1}     & 52.5         & 51.6           & 1268         & 1545   \\
IQR (ms)    & 0.317      & 0.376            & 0.646        & \textbf{0.224} & 2.67        & 2.39 
\end{tabular}
\label{tab:trecv}
\end{table}

All data were tested for normal distribution using Kolmogorov-Smirnov test \cite{kolmogorov-smirnov2008}, which returned negative results for all. Therefore, non-parametric tests are required. A Friedman test was carried out to determine if statistically significant differences appear between GALLOP, wired and wireless conditions, followed by Wilcoxon Sign-Rank Tests (with a Bonferroni correction $p<0.017$) to determine differences between conditions.  

For the position error $\epsilon$ shown in \cref{fig:eRMS} there was a statistically significant difference between communication methods, $\chi^2 = 14.9$, $p = 0.001$. Post-hoc analysis shows no significant difference between the wired and wireless conditions ($Z = -0.454$, $p = 0.65$), however there was a statistically significant reduction in $\epsilon$ between GALLOP and wired ($Z = -3.18$, $p = 0.001$) and GALLOP and wireless ($Z = -2.76$, $p = 0.006$).

Examining the velocity errors $\dot \epsilon$ shown in \cref{fig:edotRMS}, the Friedman test showed no significant difference ($\chi^2 = 2$, $p=0.368$).

For all results from the data transmission send/receive times, shown in \cref{fig:leader_times}, a statistically significant result is reported between all conditions (full results are shown in \cref{tab:Tleader-significant,tab:Fleader-significant}).

\begin{table}[htbp]
\scriptsize
\centering
\caption{Significance results for $T_{send}$ and $T_{recv}$, Leader side.}
\begin{tabular}{l|r|r|r|r|r|r}
        & \multicolumn{2}{|c|}{GALLOP/} & \multicolumn{2}{|c|}{wired/} & \multicolumn{2}{|c}{GALLOP/} \\
        & \multicolumn{2}{|c|}{wired} & \multicolumn{2}{|c|}{wireless} & \multicolumn{2}{|c}{wireless} \\
        & Send        & Receive        & Send        & Receive        & Send        & Receive        \\
            \hline
$Z$     & -150.8    & -104.5      & -112.4     & -142.7       & -143.9     & -143.9 \\
$p$     & \textless 0.001   & \textless 0.001     & \textless 0.001    & \textless 0.001      & \textless 0.001   & \textless 0.001    \\
\end{tabular}
\label{tab:Tleader-significant}
\end{table}

\begin{table}[htbp]
\scriptsize
\centering
\caption{Significance results for $T_{send}$ and $T_{recv}$, follower side.}
\begin{tabular}{l|r|r|r|r|r|r}
        & \multicolumn{2}{|c|}{GALLOP/} & \multicolumn{2}{|c|}{wired/} & \multicolumn{2}{|c}{GALLOP/} \\
        & \multicolumn{2}{|c|}{wired} & \multicolumn{2}{|c|}{wireless} & \multicolumn{2}{|c}{wireless} \\
        & Send        & Receive        & Send        & Receive        & Send        & Receive        \\
            \hline
$Z$     & -163.4    & -65.9      & -170.5     & -201.9      & -163.4     & -163.3 \\
$p$     & \textless 0.001   & \textless 0.001     & \textless 0.001    & \textless 0.001      & \textless 0.001   & \textless 0.001    \\
\end{tabular}
\label{tab:Fleader-significant}
\end{table}

Examining the results of the heuristics, for ``smoothness'' from \cref{fig:smoothness}, there was no significance between GALLOP and wired conditions ($Z=-0.791$, $p=0.429$) but significance was found between wired/wireless ($Z=-3.024$, $p=0.002$) and GALLOP/wireless ($Z=-2.83$, $p=0.005$). 

For ``responsiveness'' results shown in \cref{fig:responsive} results were similar: GALLOP/wired non-significant ($Z=-0.816$, $p=0.414$), wired/wireless and GALLOP/wireless both significant ($Z=-2.973$, $p=0.003$, $Z=-3.115$, $p=0.002$ respectively). 

Finally, this was also reflected in safety scores from \cref{fig:safe}, with GALLOP/wired showing no significant difference ($Z=-1.414$, $p=0.157$), but wired/wireless and GALLOP/wireless both significantly different ($Z=-3.217$, $p=0.001$, $Z=-3.017$, $p=0.003$ respectively). 

\section{Discussion}
Analysis of the position error results shown in \cref{fig:eRMS} shows that GALLOP produced a statistically significant reduction in error overall, with errors for wired and wireless communication very similar, and no significant differences in velocity errors shown in \cref{fig:edotRMS}. However, the heuristics scores from \cref{fig:smoothness,fig:responsive,fig:safe} show that participants perceived very little difference between GALLOP and wired communication methods, but perceived a statistically significant difference in the wireless condition. 

By examining the transmission times shown in \cref{fig:leader_times} and \cref{tab:trecv,tab:tsend,tab:Tleader-significant,tab:Fleader-significant} we can see that, in particular, the range of transmission delay for wireless communication when receiving can be very high, up to more than a second. Despite a mean value of $T_{recv}$ from \cref{tab:trecv} being reasonably low at 12.7ms, the occasional long delay can have a large affect on teleoperation performance - something that has been studied for many years \cite{sheridan1963remote,anderson1988bilateral,liu2013control,daniel1998fundamental}. In particular, performance drops significantly when delays exceeding 400ms are experienced \cite{ferrell1965delay,xu2014determination}. This accounts for the absence of a drop in performance from the mean $T_{send}$ for GALLOP of ~50ms, where the perceived delay is small enough to be compensated for by the human central nervous system. 

\section{Concluding Remarks}
Haptic teleoperation is an important application for various industries. 
This paper conducted real-world evaluation of wireless and wired technologies for haptic teleoperation. One of its key objectives was to provide a robust, responsive, and reliable wireless communication method for control commands and haptic feedback in a teleoperation system. Specifically with relation to the nuclear industry, safety and stability are particularly important for the predicted use cases.


Our results, based on objective and subjective evaluation, reveal that the use of low-power wireless control technology, based on an off-the-shelf Bluetooth 5.0 chipset, i.e.,  GALLOP, does not impact the performance of  teleoperation system. It is comparable to a standard TCP/IP wired connection, and superior to a wireless TCP/IP connection with a Wi-Fi router performing data transport. \textcolor{black}{It can be concluded that GALLOP wireless interface is a suitable low-power (and low-cost) cable replacement solution for haptic teleoperation.} 

What has been omitted from this work is investigation into other control-based techniques such as the use of wave variables \cite{niemeyer1996using,munir2001internet,aziminejad2008transparent,yang2016teleoperation}, where the usual network transport of torques and velocities (which have a \textit{multiplicative} dependence on the power-input) are transformed to wave variables (in a form where the dependence is \textit{additive}), reducing the effect of time delays on the stability and control of the system. Another method known as model-mediated control \cite{willaert2012stability,pecly2018model,achhammer2010improvement}, where the remote environment is sensed, modeled, then transported and rendered at the local controller, can also be used to improve stability in the face of time delay. Both of these methods are well documented, and could be applied using the GALLOP transport protocol, which we have plans for in the future. 

There are a number of future work directions for this work. We haven't conducted comparison against UDP which is promising for teleoperation due to (usually) lower delay and jitter. Using UDP does, however, suffers more from dropped packets. As our aims was to compare methods with similar packet loss probability, we opted to compared against TCP/IP. Future work will compare the performance of GALLOP against UDP/IP. Other areas of future work include: (a) incorporation of stability control architectures in wireless teleoperation, (b) multi-hop wireless communication based on GALLOP, and (c) the use of machine learning techniques for recovering lost packets. 


We have not compared against the user datagram protocol (UDP), which is often used for teleoperation systems due to (usually) lower delay and jitter \cite{Niemeyer1991}. Using UDP does, however, suffer more from dropped packets due to how it is implemented - we wanted to compare methods with similar packet-loss chance, so opted to compare against TCP/IP protocol. Future experiments will be carried out to compare GALLOP against UDP. 


There are many future plans for this work - we have only touched on the many configurations we would like to experiment with. For example, the GALLOP system can work in a daisy-chain network, which would extend the wireless range but introduce more complexities to the data transport. In addition, there are plans to employ an edge-intelligence system to reduce effective packet loss through the use of machine learning techniques directly in the communication layer.  


\section*{Acknowledgment}
This work was supported by UK Engineering and Physical Sciences Research Council (EPSRC No.~EP/R02572X/1) for the National Centre for Nuclear Robotics (NCNR).


\bibliographystyle{IEEEtran}
\bibliography{Ref.bib}

\end{document}